\documentclass[12pt]{article}
\usepackage{epsfig}
\def\e{\epsilon}

\def\be{\begin{equation}}
\def\ee{\end{equation}}

\def\lsim{\raise0.3ex\hbox{$<$\kern-0.75em\raise-1.1ex\hbox{$\sim$}}}
\def\gsim{\raise0.3ex\hbox{$>$\kern-0.75em\raise-1.1ex\hbox{$\sim$}}}

\def\NP{{ Nucl.\ Phys.\ }}

\def\PL{{ Phys.\ Lett.\ }}
\def\PR{{ Phys.\ Rev.\ }}

\def\PRL{{ Phys.\ Rev.\ Lett.\ }}

\def\EP{{ Eur.\ Phys.\ J.}}

\begin{document}

\thispagestyle{empty}

\noindent January 4, 2002 \hfill BI-TP 2002/01

\vskip 1.0 cm

\centerline{{\large{\bf States of Strongly Interacting
Matter}}\footnote{Invited talk at the Symposium {\sl 100 Years Werner
Heisenberg}, organized by the Alexander-von-Humboldt Foundation in
Bamberg/Germany, Sept. 26 - 30, 2001}}

\vskip 0.5cm

\centerline{\bf Helmut Satz}

\bigskip

\centerline{Fakult\"at f\"ur Physik, Universit\"at Bielefeld}
\par
\centerline{D-33501 Bielefeld, Germany}

\vskip 1cm

\noindent

\centerline{\bf Abstract:}

\medskip

I discuss the phase structure of strongly interacting matter at high 
temperatures and densities, as predicted by statistical QCD, and 
consider in particular the nature of the transition of hot hadronic
matter to a plasma of deconfined quarks and gluons.

\vskip 1cm

\noindent{\bf 1.\ Prelude}

\medskip

To speak about quark matter in a meeting dedicated to Heisenberg is
somewhat problematic. In one of his last talks, Heisenberg noted:
``There exists the conjecture that the observable hadrons consist of
non-observable quarks. But the word `consist' makes sense only if it is
possible to decompose a hadron into these quarks with an energy
expenditure much less than the rest mass of a quark" \cite{Heisenberg}.
Therefore I asked myself what arguments might have convinced Heisenberg
to revise his opinion. On philosophical level, which after all played
a significant role in Heisenberg's argumentation, one might
remember what Lucretius pointed out more than two thousand years
earlier: ``So there must be an ultimate limit to bodies, beyond
perception by our senses. This limit is without parts, is the smallest
possible thing. It can never exist by itself, but only as primordial
part of a larger body, from which no force can tear it
loose" \cite{Lucretius}. This must be one of the earliest formulations
of confinement; curiously enough, it was generally ignored by all
`atomists' before the advent of QCD. Lucretius argues that the ultimate
building blocks of matter cannot have an independent existence, since
otherwise one could ask what they are made of.

On more physical grounds, we note that the energy density of an ideal
electromagnetic plasma, consisting of electrons, positrons and photons,
is given by the Stefan-Boltzmann law
\be
\e_{\rm QED} = {\pi^2 \over 30} \left[ 2 + {7\over 8} 2 \times 2\right]
~T^4,
\label{1.1}
\ee
which counts the number of constituent species and their degrees of
freedom (two spin orientations each for electrons, positrons and
photons). For a hot and hence asymptotically free quark-gluon plasma,
the corresponding form is
\be
\e_{\rm QCD} = {\pi^2 \over 30} \left[ 2 \times 8 +
{7\over 8} 2 \times 2 \times 3 \times 3 \right] ~T^4,
\label{1.2}
\ee
which again counts the number of otherwise confined constituents and
their degrees of freedom (eight colors of gluons, three colors and
three flavors for quarks and antiquarks, and two spin orientations for
quarks and gluons). The energy density of the hot QGP thus provides
direct information on what it is made of.

A similar argument was in fact used by Zel'dovich even before the advent
of the quark model \cite{Zeldovich}. He notes that if dense stellar or
pre-stellar media should not obey the equation of state of neutron
matter, this might be an indication for other types of elementary
particles: ``It will be necessary to consider as many Fermi
distributions as there are elementary particles. The problem of the
number of elementary particles may be approached in this way, since if
some particle is in reality not elementary, it would not give rise to a
separate Fermi distribution". So the behavior of matter in the limit of
high constituent density seems to be a good way to address the question
of its ultimate building blocks.

\medskip

\noindent
{\bf 2.\ Hadronic Matter and Beyond}

\medskip

Hadrons have an intrinsic size, with a radius of about 1 fm. Hence a
hadron needs a volume $V_h=(4 \pi /3) r_h^3 \simeq 4$ fm$^3$ to exist.
This implies an upper limit $n_c$ to the density of hadronic matter,
$n_h < n_c$, with $n_c = V_h^{-1} \simeq 0.25$ fm$^{-3} \simeq 1.5~n_0$,
where $n_0 \simeq 0.17$ fm$^{-3}$ denotes standard nuclear density.
Fifty years ago, Pomeranchuk pointed out that this also leads to an
upper limit for the temperature of hadronic matter \cite{Pomeranchuk}.
An overall volume $V=NV_h$ causes the grand canonical partition function
to diverge when $T \geq T_c \simeq 1/r_h \simeq 0.2$ GeV.

This conclusion was subsequently confirmed by more detailed dynamical
accounts of hadron dynamics. Hagedorn proposed a self-similar
composition pattern for hadronic resonances, the statistical bootstrap
model, in which the degeneracy of a given resonant state is determined
by the number of ways of partitioning it into more elementary
constituents \cite{Hagedorn}. The solution of this classical
partitioning problem \cite{Euler} is a level density increasing
exponentially with mass, $\rho(m) \sim \exp\{am\}$, which leads to a
diverging partition function for an ideal resonance gas once its
temperature exceeds the value $T_H=1/a$, which turns out to be close to
the pion mass. A yet more complete and detailed description of hadron
dynamics, the dual resonance model, confirmed this exponential increase
of the resonance level density \cite{F-V,B-M}. While Hagedorn had
speculated that $T_H$ might be an upper limit of the temperature of all
matter, Cabbibo and Parisi pointed out that $T_H$ could be a critical
temperature signalling the onset of a new quark phase of strongly
interacting matter \cite{C-P}. In any case, it seems clear today that
hadron thermodynamics, based on what we know about hadron dynamics,
contains its own intrinsic limit \cite{HS-F}.

On one hand, the quark infrastructure of hadrons provides a natural
explanation of such a limit; on the other hand, it does so in a new way,
different from all previous reductionist approaches: quarks do not have
an independendent existence, and so reductionalism is at the end of the
line, in just the way proposed by Lucretius.

The limit of hadron thermodynamics can be approached in two ways. One
is by compressing cold nuclear matter, thus increasing the baryon
density beyond values of one baryon per baryon volume. The other is by
heating a meson gas to
temperatures at which collisions produce further hadrons and thus
increase the hadron density beyond values allowing each hadron its own
volume. In either case, the medium will undergo a transition from a
state in which its constituents were colorless, i.e., color-singlet
bound states of colored quarks and gluons, to a state in which the
constituents are colored. This end of hadronic matter is generally
referred to as deconfinement.

The colored constituents of deconfined matter
\begin{itemize}
\vspace*{-0.3cm}
\item{could be massive {\sl constituent quarks}, obtained if the
liberated quarks dress themselves with gluon clouds;}
\vspace*{-0.3cm}
\item{or the liberated quarks could couple pairwise to form bosonic
colored {\sl diquarks};}
\vspace*{-0.3cm}
\item{or the system could consist of unbound quarks and gluons, the
{\sl quark-gluon plasma} (QGP).}
\vspace*{-0.3cm}
\end{itemize}
One of the tasks of statistical QCD is to determine if and when these
different possible states can exist.

In an idealized world, the potential binding a heavy quark-antiquark
pair into a color-neutral hadron has the form of a string,
\be
V(r) \sim \sigma r,
\label{2.1}
\ee
where $\sigma$ specifies the string tension. For $r \to \infty$, $V(r)$
also diverges, indicating that a hadron cannot be dissociated into its
quark constituents: quarks are confined. In a hot medium, however,
thermal effects are expected to soften and eventually melt the string
at some deconfinement temperature $T_c$. This would provide the string
tension with the temperature behavior
\be
\sigma(T) = \cases{\sigma(0)~[T_c - T]^a~~&$T < T_c$,\cr
                                       0~~&$T > T_c$,}
\label{2.2}
\ee
with $a$ as critical exponent for the order parameter $\sigma(T)$.
For $T < T_c$, we then have a medium consisting of color-neutral
hadrons, for $T > T_c$ a plasma of colored quarks and gluons. The
confinement/deconfinement transition is thus the QCD version of the
insulator/conductor transition in atomic matter.

\begin{figure}[htb]
\centerline{\epsfig{file=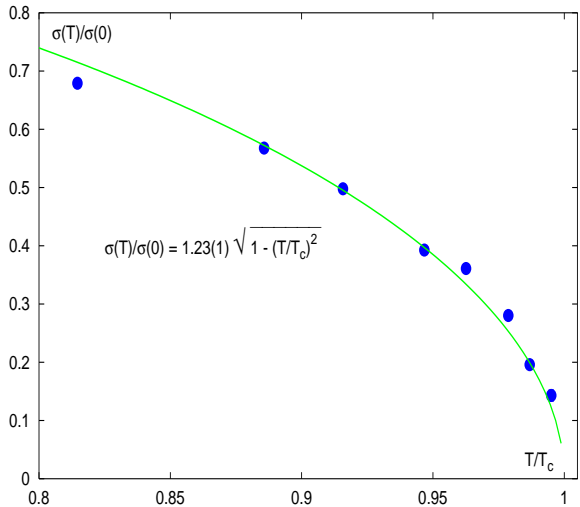,width=80mm,height=70mm}}
\caption{Temperature dependence of the string tension in $SU(3)$ 
gauge theory}
\label{F0}
\end{figure}

In the real world, the string breaks when $V(r)$ becomes larger than the
energy of two separate color singlet bound states, i.e., when the
`stretched' hadron becomes energetically more expensive than two hadrons
of normal size. It is thus possible to study the behavior of eq.\
(\ref{2.2}) only in quenched QCD, without dynamical quarks and hence
without the possibility of creating new $q \bar q$ pairs. The result 
\cite{Kacz} is shown in Fig.\ \ref{F0}, indicating that $a \simeq 0.5$. 
We shall return to the case of full QCD and string breaking in section 4.

The insulator-conductor transition in atomic matter is accompanied by
a shift in the effective constituent mass: collective effects due to
lattice oscillations, mean electron fields etc.\ give the conduction
electron a mass different from the electron mass in vacuum. In QCD, a
similar phenomenon is expected. At $T=0$, the bare quarks which make up
the hadrons `dress' themselves with gluons to form constituent quarks of
mass $M_q \simeq 300 - 350$ MeV. The mass of a nucleon then is basically
that of three constituent quarks, that of the $\rho$ meson twice $M_q$.
With increasing temperature, as the medium gets hotter, the quarks tend
to shed their dressing. In the idealized case of massless bare quarks,
the QCD Lagrangian ${\cal L}_{\rm QCD}$ possesses chiral symmetry:
four-spinors effectively reduce to two independent two-spinors. The
dynamically created constituent quark mass at low $T$ thus corresponds
to a spontaneous breaking of this chiral symmetry, and if at some high
$T=T_{\chi}$ the dressing and hence the constituent quark mass
disappears, the chiral symmetry of ${\cal L}_{\rm QCD}$ is restored.
Similar to the string tension behavior of Eq.\ (\ref{2.2}) we thus
expect
\be
M_q(T) = \cases{M_q(0)~[T_{\chi} - T]^b~~&$T < T_{\chi}$,\cr
                                       0~~&$T > T_{\chi}$.}
\label{2.3}
\ee
for the constituent quark mass: $T_{\chi}$ separates the low temperature
phase of broken chiral symmetry and the high temperature phase in which
this is restored, with $b$ as the critical exponent for the chiral
order parameter $M_q(T)$.

An obvious basic problem for statistical QCD is thus the clarification
of the relation between $T_c$ and $T_{\chi}$. In atomic physics the
electron mass shift occurs at the insulator-conductor transition; is
that also the case in QCD?

The deconfined QGP is a color conductor; what about a color
superconductor? In QED, collective effects of the medium bind electrons
into Cooper pairs, overcoming the repulsive Coulomb force between like
charges. These Cooper pairs, as bosons, condense at low temperatures
and form a superconductor. In contrast to the collective binding
effective in QED, in QCD there is already a microscopic $qq$-binding,
coupling two color triplet quarks to an antitriplet diquark. A nucleon
can thus be considered as a bound state of this diquark with the
remaining third quark,
\be
[3 \oplus 3 \oplus 3]_1 \sim [(3 \oplus 3 \to {\overline 3}) \oplus
3]_1,
\label{2.4}
\ee
leading to a color singlet state. Hence QCD provides a specific
dynamical mechanism for the formation of colored diquark bosons
and thus for color superconductivity. This possibility \cite{BL}
has created much interest and activity over the past few years
\cite{CS}.

\begin{figure}[htb]
\mbox{
\hskip0.5cm\epsfig{file=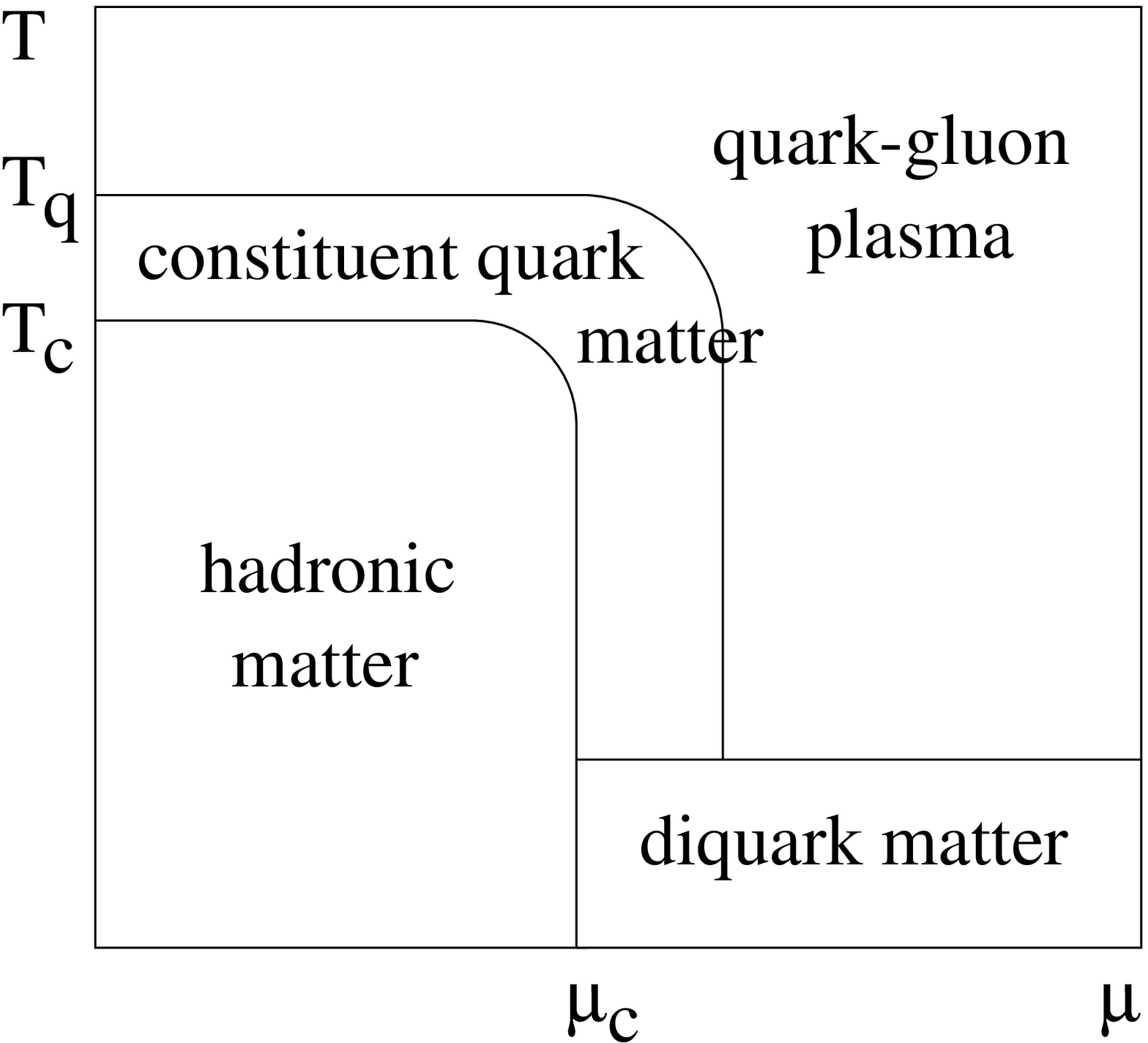,width=6cm,height=5.5cm}
\hskip2cm
\epsfig{file=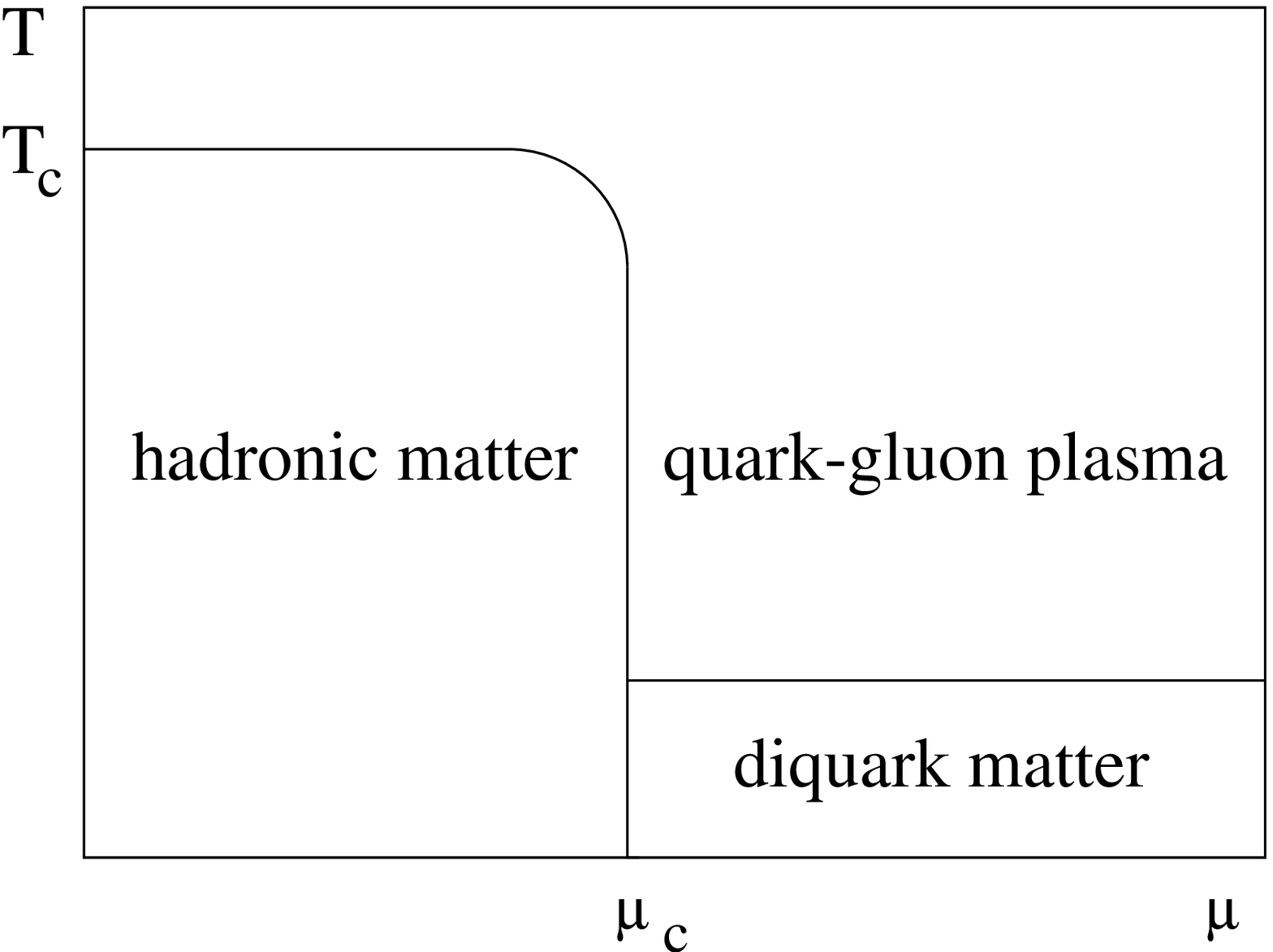,width=6cm, height=5.5cm}}
\vskip0.5cm
\vspace*{-0.5cm}
\caption{Four-phase and three-phase structure for strongly interacting
matter}
\label{F1}
\end{figure}

We thus have color deconfinement, chiral symmetry restoration and
diquark condensation as possible transitions of strongly interacting
matter for increasing temperature and/or density. This could suggest a
phase diagram of the form shown on the left
in Fig.\ \ref{F1}, with four different
phases. The results of finite temperature lattice QCD show that at least
at vanishing baryochemical potential ($\mu=0$) this is wrong, since
there deconfinement and chiral symmetry restoration coincide,
$T_c=T_{\chi}$, as the corresponding transitions in atomic physics do.
In section 4 we shall elucidate the underlying reason for this.

A second guess could thus be a three-phase diagram as shown on the right
in Fig.\
\ref{F1}, and this is in fact not in contradiction to anything so far.
In passing, we should note, however, that what we have here called the
diquark state is most likely more complex and may well consist of more
than one phase \cite{CS}.

After this conceptual introduction to the states of strongly interacting
matter, we now turn to the quantitative study of QCD at finite
temperature and vanishing baryochemical potential. In this case, along
the $\mu=0$ axis of the phase diagram \ref{F1}, the computer
simulation of lattice QCD has provided a solid quantitative basis.

\medskip

\noindent{\bf 3.\ Statistical QCD}

\medskip

The fundamental dynamics of strong interactions is defined by the QCD
Lagrangian
\be
{\cal L}_{\rm QCD} = -{1 \over 4} \left( \partial_{\mu} A_{\nu}^a -
                                       \partial_{\nu} A_{\mu}^a -
                     gf_{bc}^aA_{\mu}^bA_{\nu}^c \right)^2 -
\sum_f {\overline \psi}_{\alpha}^f \left( i\gamma_{\mu}
\partial^{\mu} + m_f -
 g \gamma_{\mu} A^{\mu} \right) \psi_{\beta}^f,
\label{3.1}
\ee
in terms of the gluon vector fields $A$ and the quark spinors $\psi$.
The corresponding thermodynamics is obtained from the partition function
\be
{\cal Z}(T,V) = \int {\cal D}A~{\cal D} \psi~{\cal D}{\overline \psi}~
\exp\{-S(A,\psi,{\overline \psi};T,V)\},
\label{3.2}
\ee
here defined as functional field integral, in which
\be
S(A,\psi,{\overline \psi};T,V) = \int_0^{1/T} d\tau \int_V d^3x
~{\cal L}_{\rm QCD}(\tau=ix_0,{\bf x})
\label{3.3}
\ee
specifies the QCD action. As usual, derivatives of $\log \cal Z$ lead to
thermodynamic observables; e.g., the temperature derivative provides the
energy density, the volume derivative the pressure of the thermal
system.

Since this system consists of interacting relativistic quantum fields,
the evaluation of the resulting expressions is highly non-trivial.
Strong interactions (no small coupling constant) and criticality
(correlations of all length scales) rule out a perturbative treatment
in the transition regions, which are of course of particular interest.
So far, the only {\sl ab initio} results are obtained through the
lattice formulation of the theory, which leads to something like a
generalized spin problem and hence can be evaluated by computer
simulation. A discussion of this approach is beyond the scope of this
survey; for an overview, see e.g. \cite{Karsch}. We shall here just
summarize the main results; it is to be noted that for computational
reasons, the lattice approach is so far viable only for vanishing
baryochemical potential, so that all results given in this section are
valid only for $\mu=0$.

As reference, it is useful to recall the energy density of an ideal gas
of massless pions of three charge states,
\be
\e_{\pi}(T) = {\pi^2 \over 30}~ 3~T^4 \simeq T^4,
\label{3.4}
\ee
to be compared to that of an ideal QGP (see Eq.\ (\ref{1.2})), which for
three massless quark flavors becomes
\be
\e_{\rm QCD}(T) \simeq 16~T^4.
\label{3.5}
\ee
The corresponding pressures are obtained through the ideal gas form
$3P(T) = \e(T)$. The main point to note is that the much larger number
of degrees of freedom of the QGP as compared to a pion gas leads at
fixed temperatures to much higher energy densities and pressures.

\begin{figure}[htb]
\mbox{
\hskip0.5cm\epsfig{file=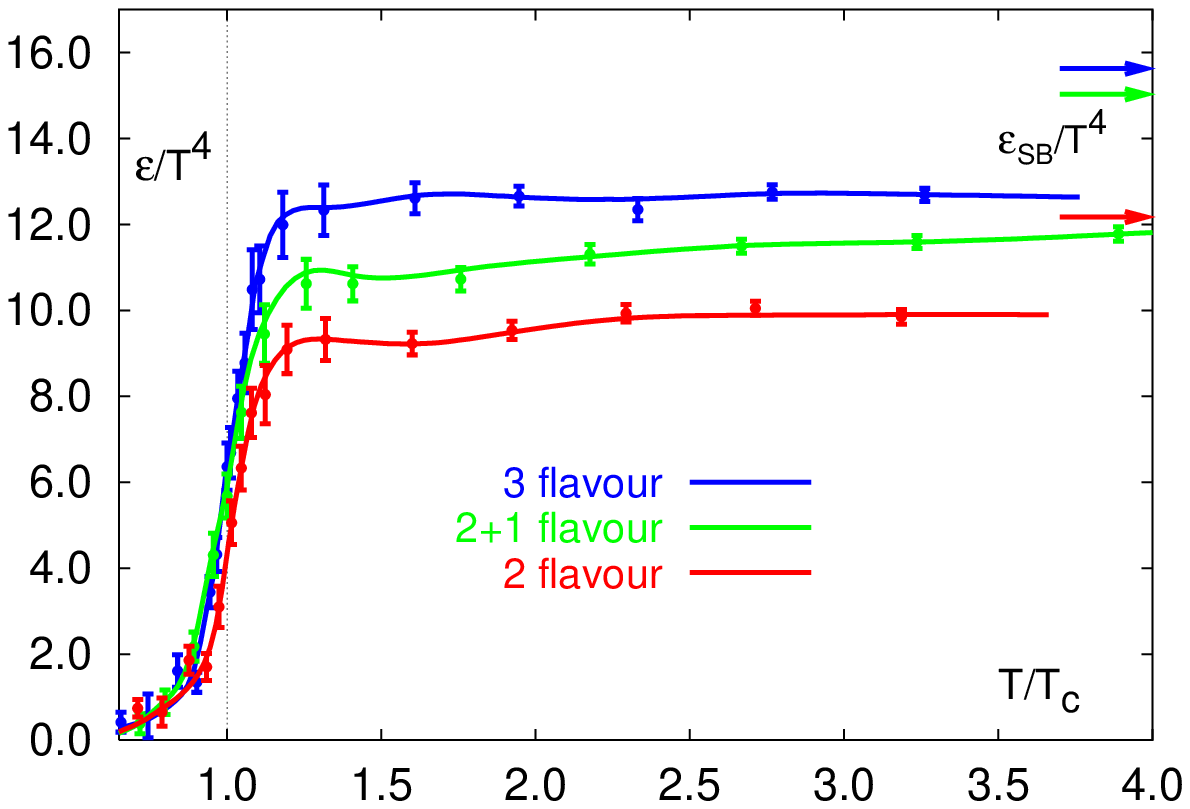,width=6.5cm,height=6cm}
\hskip1cm
\epsfig{file=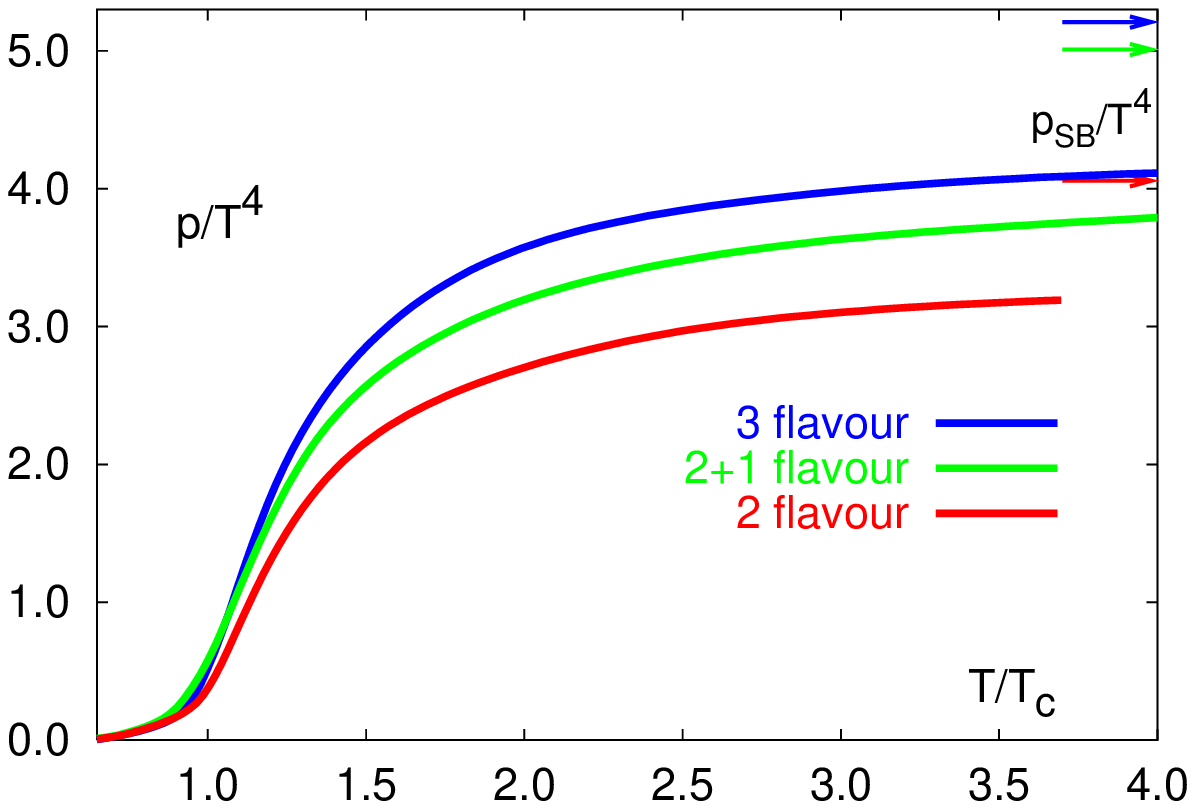,width=6.5cm, height=6cm}}
\vskip0.5cm
\vspace*{-0.5cm}
\caption{Energy density and pressure in full QCD with light dynamical
quarks}
\label{F4}
\end{figure}

The energy density and pressure have been studied in detail in finite
temperature lattice QCD with two and three light dynamical quark
species, as well as for the more realistic case of two light and
one heavier species. The results are shown in Fig.\ \ref{F4}, where it
is seen that in all cases there is a sudden increase from a state of
low to one of high values, as expected at the confinement-deconfinement
transition. To confirm the connection between the transition and the
increase of energy density or pressure, we make use of the order
parameters for deconfinement and chiral symmetry restoration; these
first have to be specified somewhat more precisely than was done in the
more conceptual discussion of section 2.

In the absence of light dynamical quarks, for $m_q \to \infty$, QCD
reduces to pure SU(3) gauge theory; the potential between two static
test quarks then has the form shown in Eq.\ (\ref{2.1}) when $T <
T_c$ and vanishes for $T \geq T_c$. The Polyakov loop expectation value
defined by
\be
\langle |L(T)| \rangle \equiv \lim_{r \to \infty} \exp\{-V(r,T)/T\} =
\cases{ 0, & confinement \cr
L(T) > 0, & deconfinement}
\label{3.6}
\ee
thus also constitutes an order parameter for the confinement state
of the medium, and it is easier to determine than the string tension
$\sigma(T)$. In lattice QCD, $L(T)$ becomes very similar to the
magnetization in spin systems; it essentially determines whether a
global $Z_3 \in SU(3)$ symmetry of the Lagrangian is present or is
spontaneously broken for a given state of the medium.

In the other extreme, for $m_q \to 0$, ${\cal L}_{\rm QCD}$ has
intrinsic chiral symmetry, and the chiral condensate $\langle \psi \bar
\psi \rangle$ provides a measure of the effective mass term in ${\cal
L}_{\rm QCD}$. Through
\be
\langle \psi \bar \psi \rangle = \cases{K(T) > 0,
& broken chiral symmetry, \cr
0, & restored chiral symmetry.}
\label{3.7}
\ee
we can determine the temperature range in which the state of the medium
shares and in which it spontaneously breaks the chiral symmetry of the
Lagrangian with $m_q=0$.

There are thus two {\sl bona fide} phase transitions in finite
temperature QCD at vanishing baryochemical potential.

For $m_q = \infty$, $L(T)$ provides a true order parameter which
specifies the temperature range $0 \leq T \leq T_c$ in which the $Z_3$
symmetry of the Lagrangian is present, implying confinement, and the
range $T > T_c$, with spontaneously broken $Z_3$ symmetry and hence
deconfinement.

For $m_q=0$, the chiral condensate defines a range $0 \leq T
\leq T_{\chi}$ in which the chiral symmetry of the Lagrangian is
spontaneously broken (quarks acquire an effective dynamical mass),
and one for $T > T_{\chi}$ in which $\langle \psi \bar \psi \rangle (T)
=0$, so that the chiral symmetry is restored. Hence here
$\langle \psi \bar \psi \rangle (T)$ is a true order parameter.

\begin{figure}[htb]
\vspace*{-2cm}
\mbox{
\epsfig{file=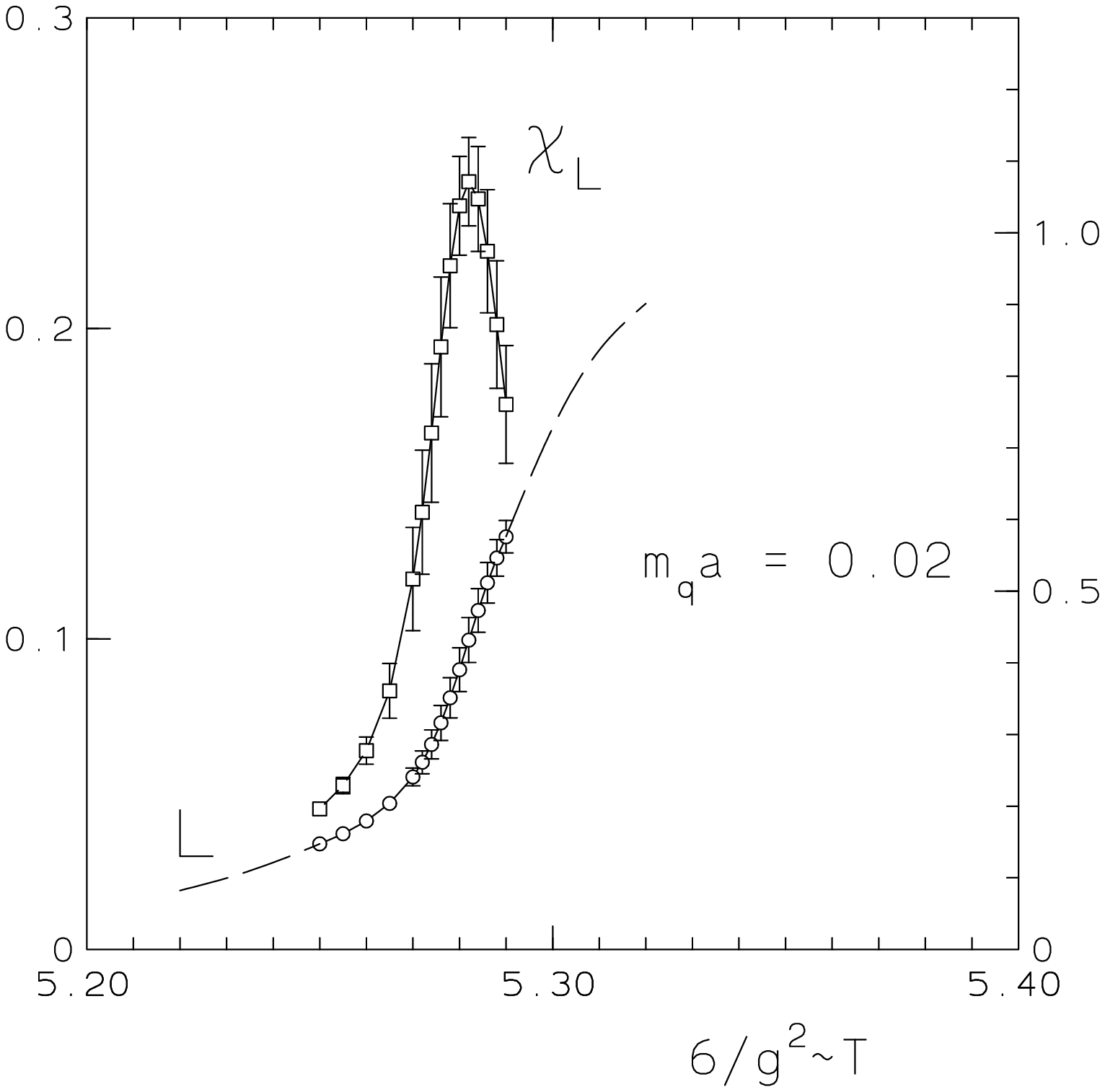,width=8cm,height=11cm}
\hskip-0.5cm
\vspace*{0.5cm}
\epsfig{file=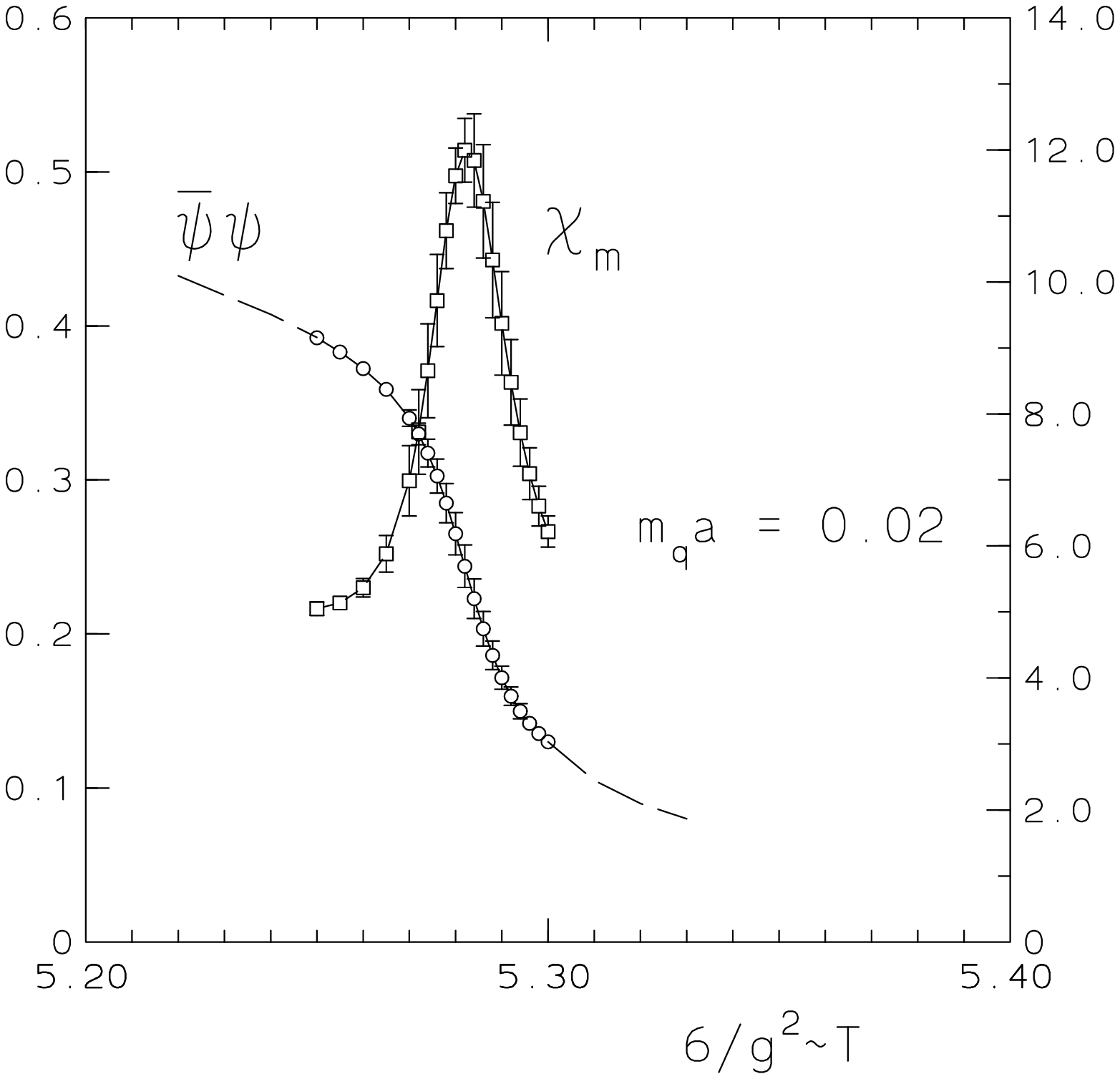,width=8cm, height=11cm}}
\vspace*{-3cm}
\caption{The temperature dependence of the Polyakov loop $L$ and 
the chiral condensate $\psi {\bar \psi}$, as well as of the 
corresponding susceptibilities.}
\label{F5}
\end{figure}

In the real world, the (light) quark mass is small but finite: $0 < m_q
< \infty$. This means that the string breaks for all temperatures, even
for $T=0$, so that $L(T)$ never vanishes. On the other hand, with $m_q
\not= 0$, the chiral symmetry of ${\cal L}_{\rm QCD}$ is explicitly
broken, so that $\langle \psi \bar \psi \rangle$ never vanishes. It is
thus not clear if some form of critical behavior remains, and we are
therefore confronted by two basic questions:
\begin{itemize}
\vspace*{-0.2cm}
\item{how do $L(T)$ and $\langle \psi \bar \psi \rangle (T)$ behave for
small but finite $m_q$? Is it still possible to identify transition
points, and if so,}
\vspace*{-0.2cm}
\item{what if any relation exists between $T_c$ and $T_{\chi}$?}
\vspace*{-0.2cm}
\end{itemize}
In Fig.\ \ref{F5} we show the lattice results for two light quark
species; it is seen that $L(T)$ as well as $\langle \psi \bar \psi
\rangle (T)$ still experience very strong variations, so that clear 
transition temperatures can be identified through the peaks in the
corresponding susceptibilities, also shown in the figure. Moreover, 
the two peaks occur at the same temperature; one thus finds here
(and in fact for all small values of $m_q$) that $T_c =T_{\chi}$, so 
that the two `quasi-critical' transitions of deconfinement and chiral 
symmetry restoration coincide.

Although all lattice calculations are performed for non-vanishing bare
quark mass in the Lagrangian, results obtained with different $m_q$
values can be extrapolated to the chiral limit $m_q=0$. The resulting
transition temperatures are found to be $T_c(N_f=2) \simeq 175$ MeV and
$T_c(N_f=3) \simeq 155$ MeV for two and three light quark flavors,
respectively. The order of the transition is still not fully determined.
For $N_f=3$ light quark species, one obtains a first order transition.
For two light flavors, a second order transition is predicted \cite{P-W},
but not yet unambiguously established.

\medskip

\noindent{\bf 4.\ The Nature of Deconfinement}

\medskip

In this last section I want to consider in some more detail two basic
aspects which came up in the previous discussion of deconfinement:
\begin{itemize}
\vspace*{-0.2cm}
\item{Why do deconfinement and chiral symmetry restoration coincide for
all (small) values of the input quark mass?}
\vspace*{-0.2cm}
\item{Is there still some form of critical behavior when $m_q \not= 0$?}
\vspace*{-0.2cm}
\end{itemize}
Both features have recently been addressed, leading to some first and
still somewhat speculative conclusions which could, however, be more
firmly established by further lattice studies.

In the confined phase of pure gauge theory, we have $L(T)=0$, the Polyakov
loop as generalized spin is disordered, so that the state of the system
shares the $Z_3$ symmetry of the Lagrangian. Deconfinement then corresponds
to ordering through spontaneous breaking of this $Z_3$ symmetry, making
$L\not=0$. In going to full QCD, the introduction of dynamical quarks 
effectively brings in an external field $H(m_q)$, which in principle
could order $L$ in a temperature range where it was previously disordered.

\begin{figure}[htb]
\centerline{\epsfig{file=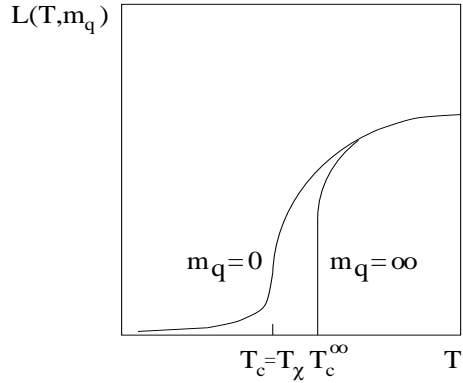,width=60mm,height=50mm}}
\caption{Temperature dependence of the Polyakov loop in the chiral
and the pure gauge theory limits} 
\label{F6}
\end{figure}

Since $H \to 0$ for $m_q \to \infty$, $H$ must for large quark masses be
inversely proportional to $m_q$. On the other hand, since $L(T)$ shows
a rapid variation signalling an onset of deconfinement even in the chiral
limit, the relation between $H$ and $m_q$ must be different for $m_q \to 0$.
We therefore conjecture \cite{G-O,DLS} that $H$ is determined by the 
effective constituent quark mass $M_q$, setting
\be
H \sim {1 \over m_q + c \langle \psi {\bar \psi} \rangle},
\label{4.1}
\ee
since the value of $M_q$ is determined by the amount of chiral symmetry
breaking and hence by the chiral condensate. From Eq.\ (\ref{4.1}) we 
obtain
\begin{itemize}
\vspace*{-0.2cm}
\item{for $m_q \to \infty$, $H \to 0$, so that we recover the pure gauge
theory limit;} 
\vspace*{-0.2cm}
\item{for $m_q \to 0$, we have
\be
\langle \psi {\bar \psi} \rangle = \cases{
{\rm large},~H~{\rm small}, L~ {\rm disordered}, 
& for $T \leq T_{\chi}$;\cr
{\rm small},~H~{\rm large}, L~{\rm ordered}, & for $T > T_{\chi}$.}
\ee}
\vspace*{-0.3cm}
\end{itemize}
\vskip -0.3cm
In full QCD, it is thus the onset of chiral symmetry restoration that drives
the onset of deconfinement, by ordering the Polyakov loop at a temperature
value below the point of spontaneous symmetry breaking \cite{DLS}.
In Fig.\ \ref{F6} we 
compare the behavior of $L(T)$ in pure gauge theory to that in the chiral 
limit of QCD. In both cases, we have a rapid variation at some temperature
$T_c$. This variation is for $m_Q \to \infty$ due to the spontaneous breaking
of the $Z_3$ symmetry of the Lagrangian at $T=T_c^{\infty}$; for $m_q \to 0$, 
the Lagrangian
retains at low temperatures an approximate $Z_3$ symmetry which is explicitly
broken at $T_{\chi}$ by an external field which becomes strong when the
chiral condensate vanishes. For this reason, the peaks in the Polyakov loop 
and the chiral susceptibility coincide and we have $T_{\chi}=T_c < 
T_c^{\infty}$. 
 
A quantitative test of this picture can be obtained from finite temperature
lattice QCD. It is clear that in the chiral limit $m_q \to 0$, the chiral
susceptibilities (derivatives of the chiral condensate
$\langle \psi {\bar \psi} \rangle$) will diverge at $T=T_{\chi}$. If 
deconfinement is indeed driven by chiral symmetry restoration, i.e., if 
$L(T,m_q)=L(H(T),m_q)$ with $H(T)=H(\langle \psi {\bar \psi} \rangle(T))$ 
as given in Eq.\ (\ref{4.1}), than also the Polyakov loop susceptibilities 
(derivatives of $L$) must diverge in the chiral limit. Moreover, these 
divergences must be governed by the critical exponents of the chiral
transition.

\begin{figure}[thb]
\mbox{
\epsfig{file=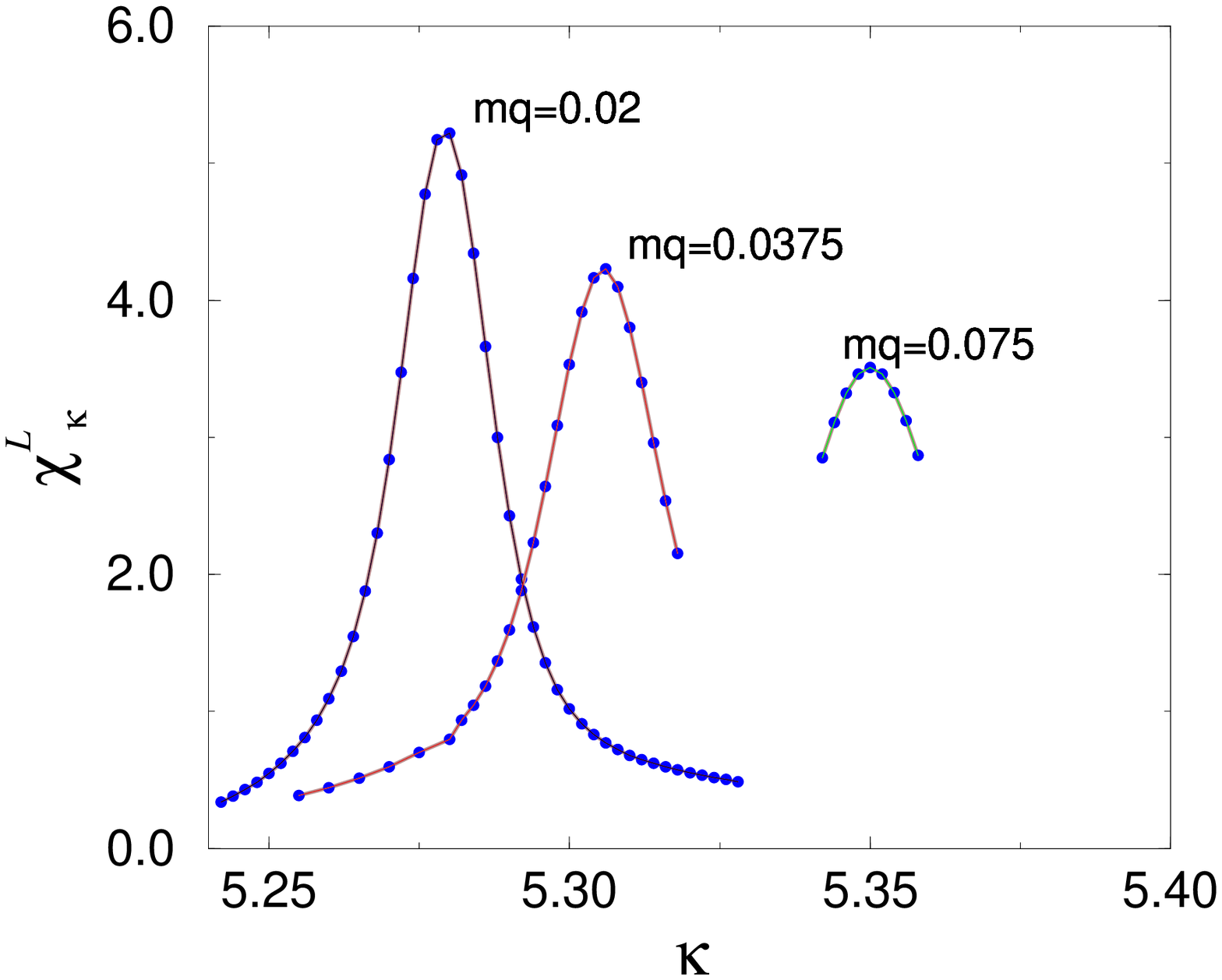,width=7cm,height=7cm}
\hskip0.5cm
\epsfig{file=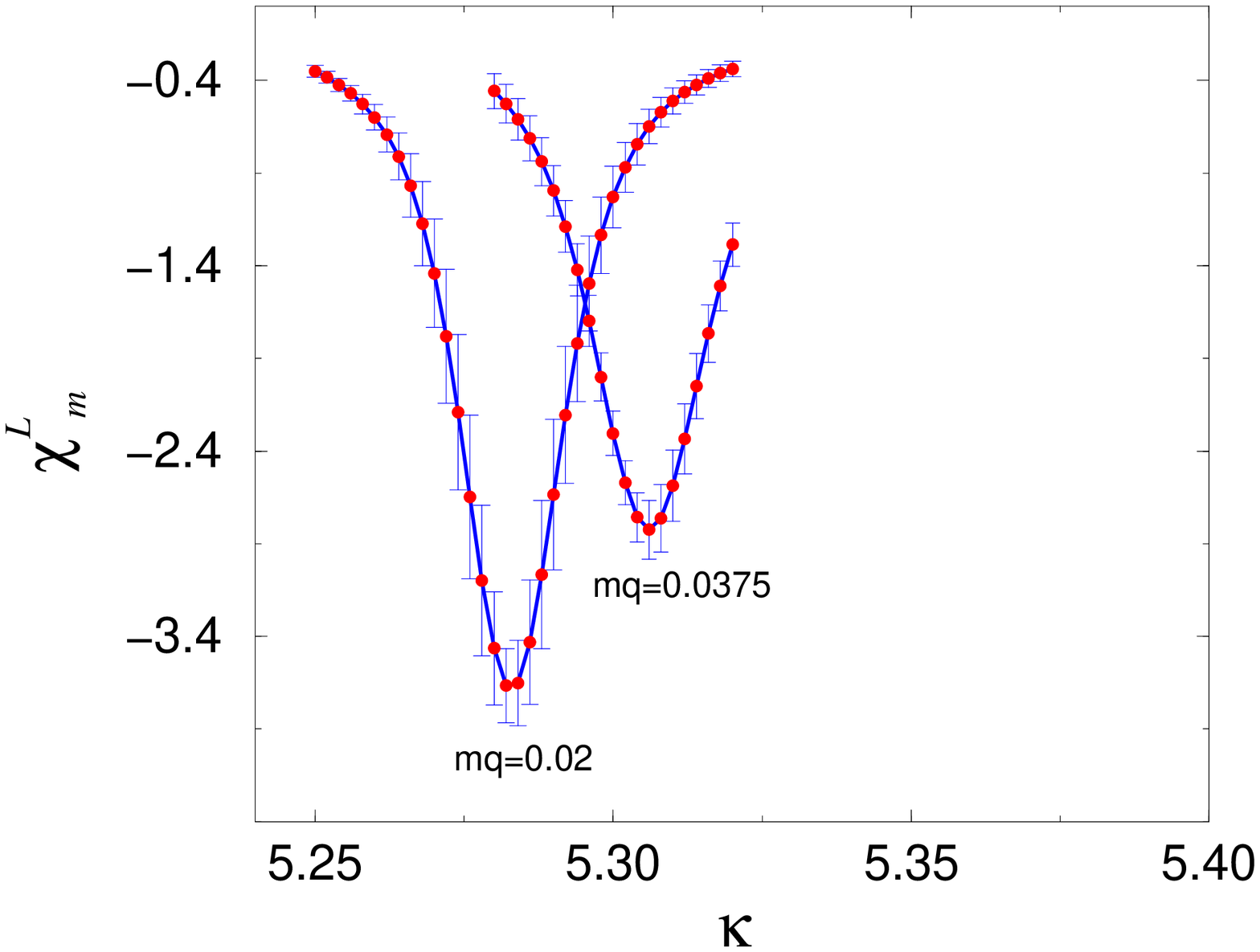,width=7cm, height=7cm}}
\caption{The Polyakov loop susceptibilities with respect to temperature
$\chi_{\kappa}^L$ (left) and to quark mass $\chi_m^L$ (right) as function 
of the temperature variable $\kappa=6/g^2$ for different quark massses.} 
\label{F7}
\end{figure}

Preliminary lattice studies support our picture \cite{DLS}. In Fig.\ \ref{F7}
we see that the peaks in the Polyakov loop susceptibilities as function of 
the effective temperature increases as $m_q$ decreases, suggesting
divergences in the chiral limit. Further lattice calculations for smaller
$m_q$ (which requires larger lattices) would certainly be helpful. The
question of critical exponents remains so far completely open, even for
the chiral condensate and its susceptibilities. 

\begin{figure}[htb]
\mbox{
\hskip1.5cm\epsfig{file=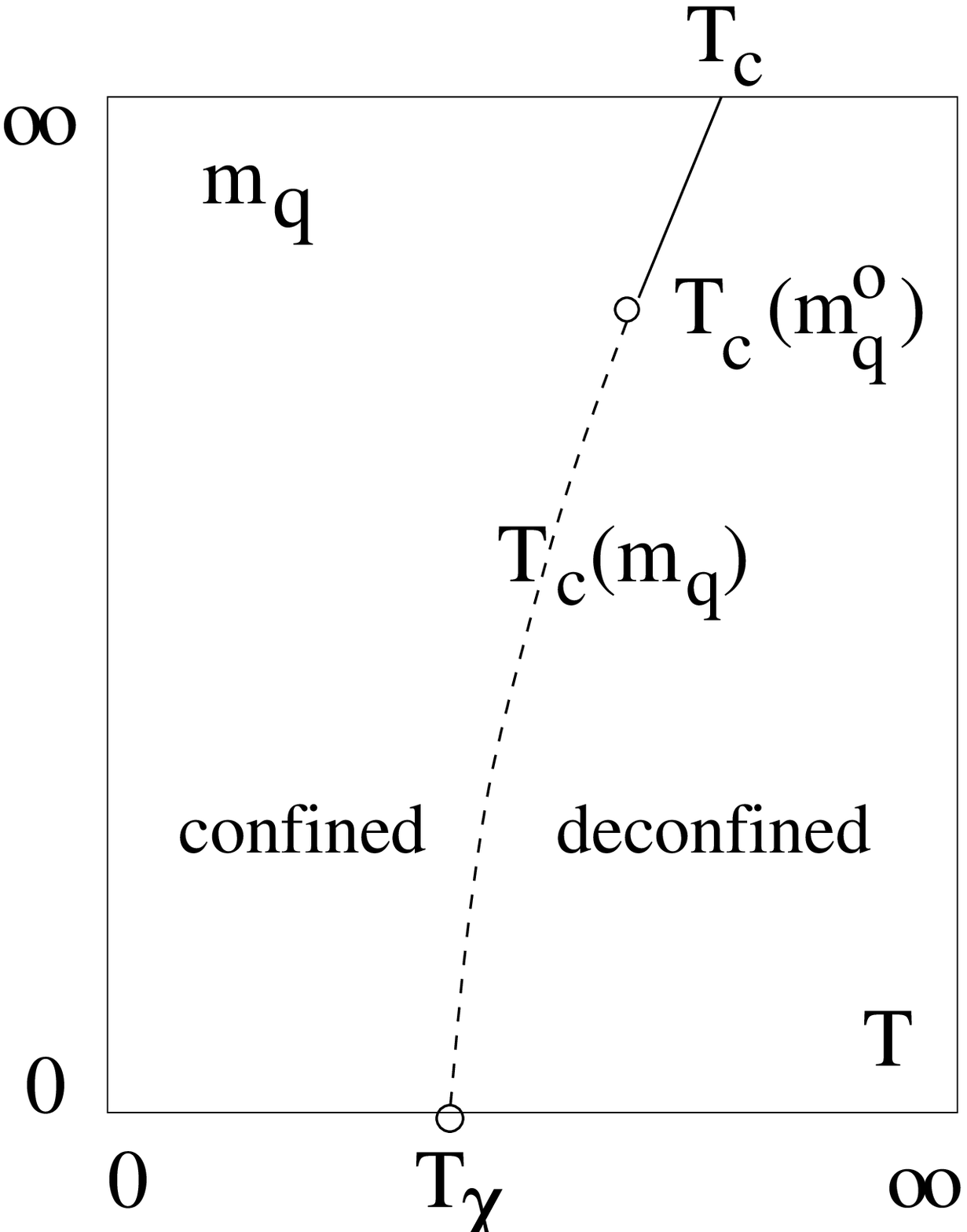,width=5cm,height=6cm}
\hskip3cm
\epsfig{file=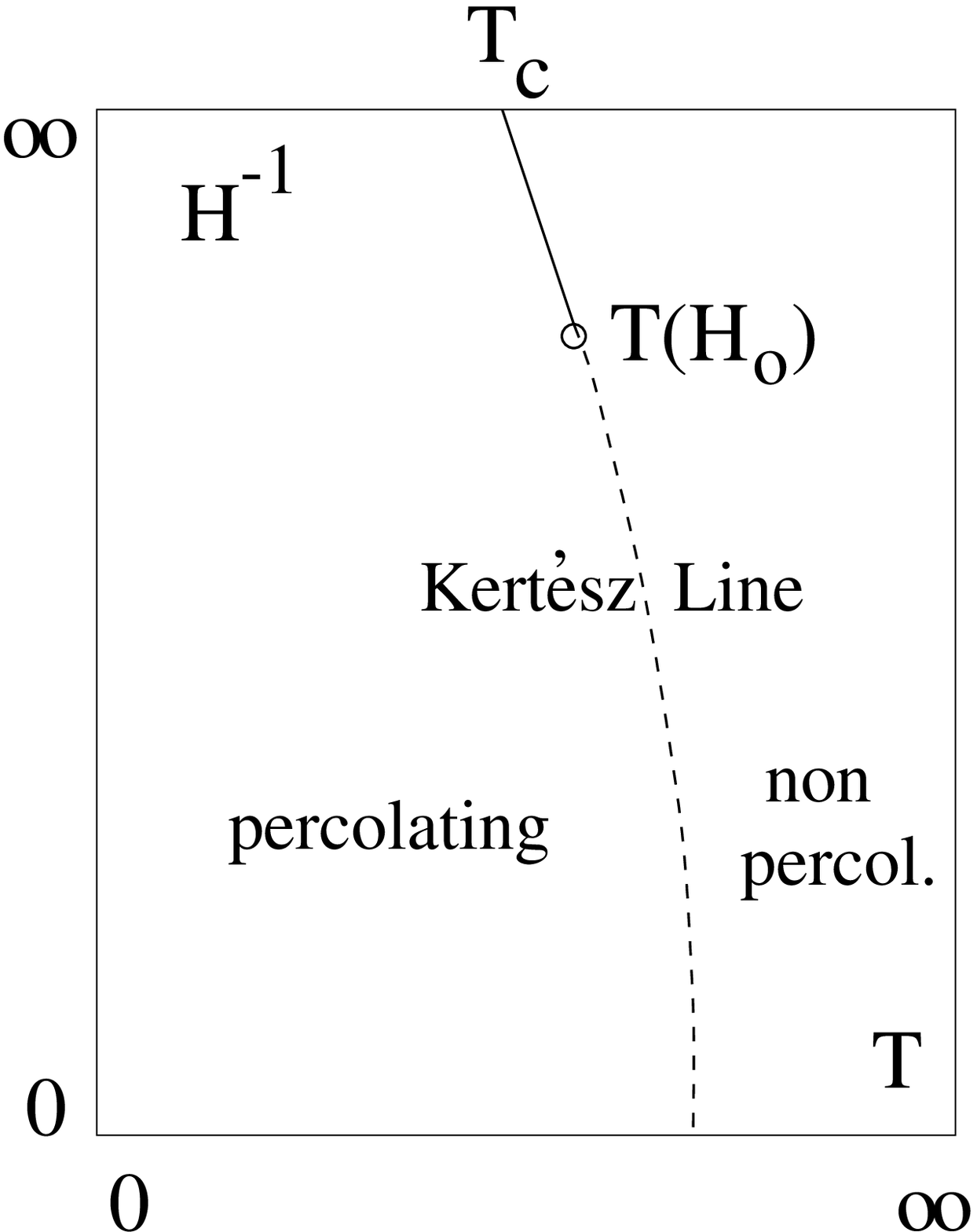,width=5cm, height=6cm}}
\vskip0.5cm
\caption{The phase structure of QCD (left) and of the 3-d 3-state 
Potts model (right)}
\label{F8}
\end{figure}

Next we want to consider the nature of the transition for $0 < m_q < \infty$.
For finite quark mass neither the Polyakov loop nor the chiral condensate 
constitute genuine order parameters, since both are non-zero at all finite 
temperatures. Is there then any critical behavior? For pure $SU(3)$ gauge 
theory, the deconfinement transition is of first order, and the associated
discontinuity in $L(T)$ at $T_c$ cannot disappear immediately for $m_q < 
\infty$. Hence in a  certain mass range $m_q^0 < m_q \leq \infty$, 
a discontinuity in $L(T)$ remains; it vanishes for $m_q^0$ at the endpoint
$T_c(m_q^0)$ in the $T-m_q$ plane; see Fig.\ \ref{F8}. For $m_q=0$,
we have the genuine chiral transition (perhaps of second order \cite{P-W})
 at $T_{\chi}$, which, as we just saw, 
leads to critical behavior also for the Polyakov loop, so that here 
$T_c(m_q=0)=T_{\chi}$ is a true critical temperature. What happens between 
$T_c(m_q^0)$ and $T_c(m_q=0)=T_{\chi}$? The dashed line in Fig.\ \ref{F8}
separating the hadronic phase from the quark-gluon plasma is not easy to 
define unambiguously: it could be obtained from the peak position of chiral 
and/or Polyakov loop susceptibilities \cite{pseudo}, or from maximizing 
the correlation length in the medium \cite{caselle}. In any case, it does 
not appear to be related to thermal critical behavior in a strict 
mathematical sense.

An interesting new approach to the behavior along this line could be
provided by cluster percolation \cite{S-A}. For spin systems without
external field, the thermal magnetization transition can be equivalently
described as a percolation transition of suitably defined clusters
\cite{F-K,C-K}. We recall that a system is said to percolate once the
size of clusters reach the size of the system (in the infinite volume
limit). One can thus characterize the Curie point of a spin system 
either as the point where with decreasing temperature spontaneous symmetry 
breaking sets in, or as the point where the size of suitably bonded 
like-spin clusters diverges: the critical indices of the percolation
transition are identical to those of the magnetization transition.

For non-vanishing external field $H$, 
there is no more thermal critical behavior; for the 2d Ising model, as
illustration, the partition function now is analytic. In a
purely geometric description, however, the percolation transition persists 
for all $H$, but the critical indices now are those of random percolation
and hence differ from the thermal (magnetization) indices.
For the 3d three state Potts' model (which also has a first order
magnetization transition),
the resulting phase diagram is shown on the right of Fig.\ \ref{F8};
here the dashed line, the so-called Kert\'esz line \cite{K}, is defined
as the line of the geometric critical behavior obtained from cluster
percolation. The phase on the low temperature side of the Kert\'esz line
contains percolating clusters, the high temperature phase does not
\cite{Potts}.
Comparing this result to the $T-m_q$ diagram of QCD, one is tempted
to speculate that deconfinement for $ 0 < m_q < \infty$ corresponds to 
the Kert\'esz line of QCD \cite{HS}. First studies have shown that in
pure gauge theory, one can in fact describe deconfinement through
Polyakov loop percolation \cite{F-S,S}. It will indeed be interesting
to see if this can be extended to full QCD.

\medskip

\noindent{\bf 5.\ Summary}

\medskip

We have seen that at high temperatures and vanishing baryon density,
hadronic matter becomes a plasma of deconfined colored quarks and 
gluons. In contrast, at high baryon densities and low temperatures,
one expects a condensate of colored diquarks. The quark-gluon plasma
constitutes the conducting, the diquark condensate the superconducting
phase of QCD. 

For vanishing baryon density, the deconfinement transition has been
studied extensively in finite temperature lattice QCD. In pure $SU(N)$
gauge theory (QCD for $m_q \to \infty$), deconfinement is due to the
spontaneous breaking of a globel $Z_N$ symmetry of the Lagrangian and
structurally of the same nature as the magnetization transition in 
$Z_N$ spin systems. In full QCD, deconfinement is triggered by a strong
explicit breaking of the $Z_N$ symmetry through an external field induced
by the chiral condensate $\langle \psi {\bar \psi} \rangle$. Hence for
$m_q=0$ deconfinement coincides with chiral symmetry restoration.

For finite quark mass, $0,<,m_q,<,\infty$, it does not seem possible to
define thermal critical behavior in QCD. On the other hand, spin systems
under similar conditions retain geometric cluster percolation as a
form of critical behavior even when there is no more thermal criticality.
It is thus tempting to speculate that cluster percolation will allow
a definition of color deconfinement in full QCD as genuine but 
geometric critical behavior. 

\vskip1cm

\noindent
{\bf Acknowledgements}

It is a pleasure to thank S.\ Digal, J.\ Engels, S.\ Fortunato, 
F.\ Karsch and E.\ Laermann for many helpful discussions. The financial 
support of the German Ministry of Science (contract 06BI902) and of the 
GSI Darmstadt (contract BI-SAT) is gratefully acknowledged.

\end{document}